\documentclass[a4paper]{article}
\usepackage{graphicx}
\usepackage{amsmath}

\begin{document}

\title{Perfect A/D conversion of entanglement}
\author{Xiao-Yu Chen\\
{\small \ China Institute of Metrology, Hangzhou 310018,China;}}
\date{}
\maketitle

\begin{abstract}
We investigate how entanglement can be perfectly transfered between
continuous variable and qubits system. We find that a two-mode squeezed
vacuum state can be converted to the product state of an infinitive number
of two-qubit states while keeping the entanglement. The reverse process is
also possible. The interaction Hamitonian is a kind of non-linear
Jaynes-Cumings Hamiltonian.


\end{abstract}

Quantum information processing (QIP) has been extensively studied for a
qubit system which is a quantum extension of a bit, spanning two-dimensional
Hilbert space. A qubit is realized by a electronic spin, a two-level atom,
the polarization of a photon and a superconductor among others. Parallelly,
much attentions have been paid to the QIP of quantum continuous variable
(CV) system which is a quantum extension of analog information in classical
information theory. CV physical systems such as a harmonic oscillator, a
rotator and a light field are defined in infinitive-dimensional Hilbert
space. While conversions of analog to digital (A/D) and digital to analog
(D/A) are quite usual in information processing, qubit and CV systems are
nearly always treated separately. There have been some pilot works on how to
entangle two separate qubits by an entangled Gaussian field, the efficient
of the transfer is not high. We would propose a scheme of perfect
transferring the entanglement in this letter.

The two-mode squeezed vacuum state $\left| \Psi \right\rangle _{AB}=\sqrt{%
1-\lambda ^2}\sum_{m=0}^\infty \lambda ^m\left| m\right\rangle _A\otimes
\left| m\right\rangle _B$, where $\lambda =\tanh r$ with $r$ the squeezing
parameter. The entanglement of the state is $E(\left| \Psi \right\rangle
)=\cosh ^2r\log \cosh ^2r-$ $\sinh ^2r\log \sinh ^2r$. The interaction
between different systems can cause the transfer of entanglement between the
systems. The scheme of the system considered is that two individual qubits
each interacting with one entangled part of the field. The whole system will
evolve in the way of $\ U\left( t\right) \rho _{AB}\left( 0\right) \otimes
\rho _{CD}\left( 0\right) U^{+}\left( t\right) $ , where $U\left( t\right)
=\exp [-\frac i\hbar (H_{AC}+H_{BD})t]$ is the evolution operator in
interaction picture, and $\rho _{AB}\left( 0\right) =$ $\left| \Psi
\right\rangle _{AB\text{ }AB}\left\langle \Psi \right| $is the initial state
of the CV system while $\rho _{CD}\left( 0\right) =\left| -\right\rangle _{C%
\text{ }C}\left\langle -\right| _C\otimes \left| -\right\rangle _{D\text{ }%
D}\left\langle -\right| $ is the initial state of the qubit system. Firstly
suppose the model Hamiltonian of entanglement transfer from CV system to
qubit system or vice versa is
\begin{equation}
H_1=\hbar \Omega \left( \sqrt{n}a^{+}\sigma _{-}+a\sqrt{n}\sigma _{+}\right)
,  \label{wave1}
\end{equation}
where $a$ and $a^{+}$ are the photon annihilation and generation operators
respectively, $n=a^{+}a$, $\sigma _{-}$ and $\sigma _{+}$ are operators
which convert the atom form its excited state $\left| +\right\rangle $ to
ground state $\left| -\right\rangle $ and from ground state to excited state
respectively. The Hamiltonian (\ref{wave1}) can be considered as a kind of
nonlinear Jaynes-Cummings model\cite{Mista}. Then $\exp [-\frac i\hbar
H_1t_1]\left| m,-\right\rangle =\cos (m\Omega t_1)\left| m,-\right\rangle
-i\sin (m\Omega t_1)\left| m-1,+\right\rangle .$ If the interaction time $%
t_1 $ is adjusted in such a way that $\Omega t_1=\pi /2$ then $\exp [-\frac
i\hbar H_1t_1]\left| 2m,-\right\rangle =(-1)^m\left| 2m,-\right\rangle $ and
$\exp [-\frac i\hbar H_1t_1]\left| 2m+1,-\right\rangle =-i(-1)^m\left|
2m,+\right\rangle .$ Apply the evolution operator $U_1(t_1)=\exp [-\frac
i\hbar (H_{1AC}+H_{1BD})t_1]$ to the state$\left| \Psi \right\rangle
_{AB}\left| --\right\rangle _{CD\text{ }}^{(1)}$ , then
\begin{equation}
U_1(t_1)\left| \Psi \right\rangle _{AB}\left| --\right\rangle _{CD\text{ }%
}^{(1)}=\left| \Psi \right\rangle _{AB}^{(1)}\left| \Phi \right\rangle
_{CD}^{(1)}
\end{equation}
with $\left| \Psi \right\rangle _{AB}^{(1)}=\sqrt{1-\lambda ^4}%
\sum_{m=0}^\infty \lambda ^{2m}\left| 2m\right\rangle _A\left|
2m\right\rangle _B$ and $\left| \Phi \right\rangle _{CD}^{(1)}=\frac 1{\sqrt{%
1+\lambda ^2}}(\left| --\right\rangle _{CD}^{(1)}-\lambda \left|
++\right\rangle _{CD}^{(1)}).$ It should be noticed that the state after
evolution is a product state of CV system state and two qubit state. The CV
state $\left| \Psi \right\rangle _{AB}^{(1)}$ has even number of photons in
each mode. We can separate the two qubit state $\left| \Phi \right\rangle
_{CD}^{(1)}$ from the combined state, then append another vacuum two qubit
state $\left| --\right\rangle _{CD}^{(2)}$of $CD$ partite to state $\left|
\Psi \right\rangle _{AB}^{(1)}$ , the new state will be $\left| \Psi
\right\rangle _{AB}^{(1)}\left| --\right\rangle _{CD\text{ }}^{(2)}.$ We
would design another interaction Hamiltonian to assign the entanglement of
CV state to two qubit state. The Hamiltonian will be $H_2=\hbar \Omega
\left( \sqrt{n}a^{+}\frac 1{\sqrt{n}}a^{+}\sigma _{-}+a\frac 1{\sqrt{n}}a%
\sqrt{n}\sigma _{+}\right) ,$the evolution will be $U_2(t_2)\left| \Psi
\right\rangle _{AB}^{(1)}\left| --\right\rangle _{CD\text{ }}^{(2)}=\left|
\Psi \right\rangle _{AB}^{(2)}\left| \Phi \right\rangle _{CD}^{(2)}$ with
the interaction time $t_2$ $=\pi /(4\Omega )$, and $\left| \Psi
\right\rangle _{AB}^{(2)}=\sqrt{1-\lambda ^8}\sum_{m=0}^\infty \lambda
^{4m}\left| 4m\right\rangle _A\left| 4m\right\rangle _B$ , $\left| \Phi
\right\rangle _{CD}^{(2)}=\frac 1{\sqrt{1+\lambda ^4}}(\left|
--\right\rangle _{CD}^{(2)}-\lambda ^2\left| ++\right\rangle _{CD}^{(2)}).$
Then we move from the second two qubit to the vacuum state of the third two
qubit of $CD$ partite and so on. The $k-th$ Hamiltonian will be $H_k=\hbar
\Omega [n(\frac 1{\sqrt{n}}a^{+})^{2^{k-1}}\sigma _{-}+(a\frac 1{\sqrt{n}%
})^{2^{k-1}}n\sigma _{+}]$ and interaction time $t_k=\pi /(2^k\Omega ).$ The
whole state will be
\begin{equation}
U_k(t_k)\cdots U_2(t_2)U_1(t_1)\left| \Psi \right\rangle _{AB}(\left|
--\right\rangle _{\text{ }}^{(1)}\left| --\right\rangle ^{(2)}\cdots \left|
--\right\rangle ^{(k)})_{CD}=\left| \Psi \right\rangle _{AB}^{(k)}(\left|
\Phi \right\rangle _{\text{ }}^{(1)}\left| \Phi \right\rangle ^{(2)}\cdots
\left| \Phi \right\rangle ^{(k)})_{CD},
\end{equation}
with $\left| \Psi \right\rangle _{AB}^{(k)}=\sqrt{1-\lambda ^{2^{k+1}}}%
\sum_{m=0}^\infty \lambda ^{2^km}\left| 2^km\right\rangle _A\left|
2^km\right\rangle _B$ , $\left| \Phi \right\rangle _{CD}^{(k)}=\frac 1{\sqrt{%
1+\lambda ^{2^k}}}(\left| --\right\rangle _{CD}^{(k)}-\lambda
^{2^{k-1}}\left| ++\right\rangle _{CD}^{(k)}).$ The entanglement transferred
to qubits system is
\begin{eqnarray}
E(\prod_{j=1}^k\left| \Phi \right\rangle _{CD\text{ }}^{(j)})
&=&\sum_{j=1}^kE(\left| \Phi \right\rangle _{CD\text{ }}^{(j)})=%
\sum_{j=1}^k[\log (1+\lambda ^{2^j})-\frac{\lambda ^{2^j}}{1+\lambda ^{2^j}}%
2^j\log \lambda ] \\
&=&\log \frac{1-\lambda ^{2^{k+1}}}{1-\lambda ^2}-(\frac{\lambda ^2}{%
1-\lambda ^2}-\frac{2^k\lambda ^{2^{k+1}}}{1-\lambda ^{2^{k+1}}})\log
\lambda ^2.  \nonumber
\end{eqnarray}
The entanglement remained at the CV system is $E(\left| \Psi \right\rangle
_{AB}^{(k)})=-\log (1-\lambda ^{2^{k+1}})-\frac{2^{k+1}\lambda ^{2^{k+1}}}{%
1-\lambda ^{2^{k+1}}}\log \lambda .$The total entanglement remains unchanged
for each $k$, $E(\prod_{j=1}^k\left| \Phi \right\rangle _{CD\text{ }%
}^{(j)})+E(\left| \Psi \right\rangle _{AB}^{(k)})=E(\left| \Psi
\right\rangle _{AB}).$When $k\rightarrow \infty ,$ $\lambda
^{2^{k+1}}\rightarrow 0$, thus $E(\left| \Psi \right\rangle
_{AB}^{(k)})\rightarrow 0,$ the entanglement transferred to the qubit system
tends to $E(\left| \Psi \right\rangle _{AB})$. The entanglement is perfectly
transferred. The entanglement transfer is depicted in Fig. (1) for different
value of receiving qubit pair number $k$.
\begin{figure}[tbp]
\includegraphics[ trim=0.000000in 0.000000in -0.138042in 0.000000in,
height=2.0081in, width=2.5097in ]{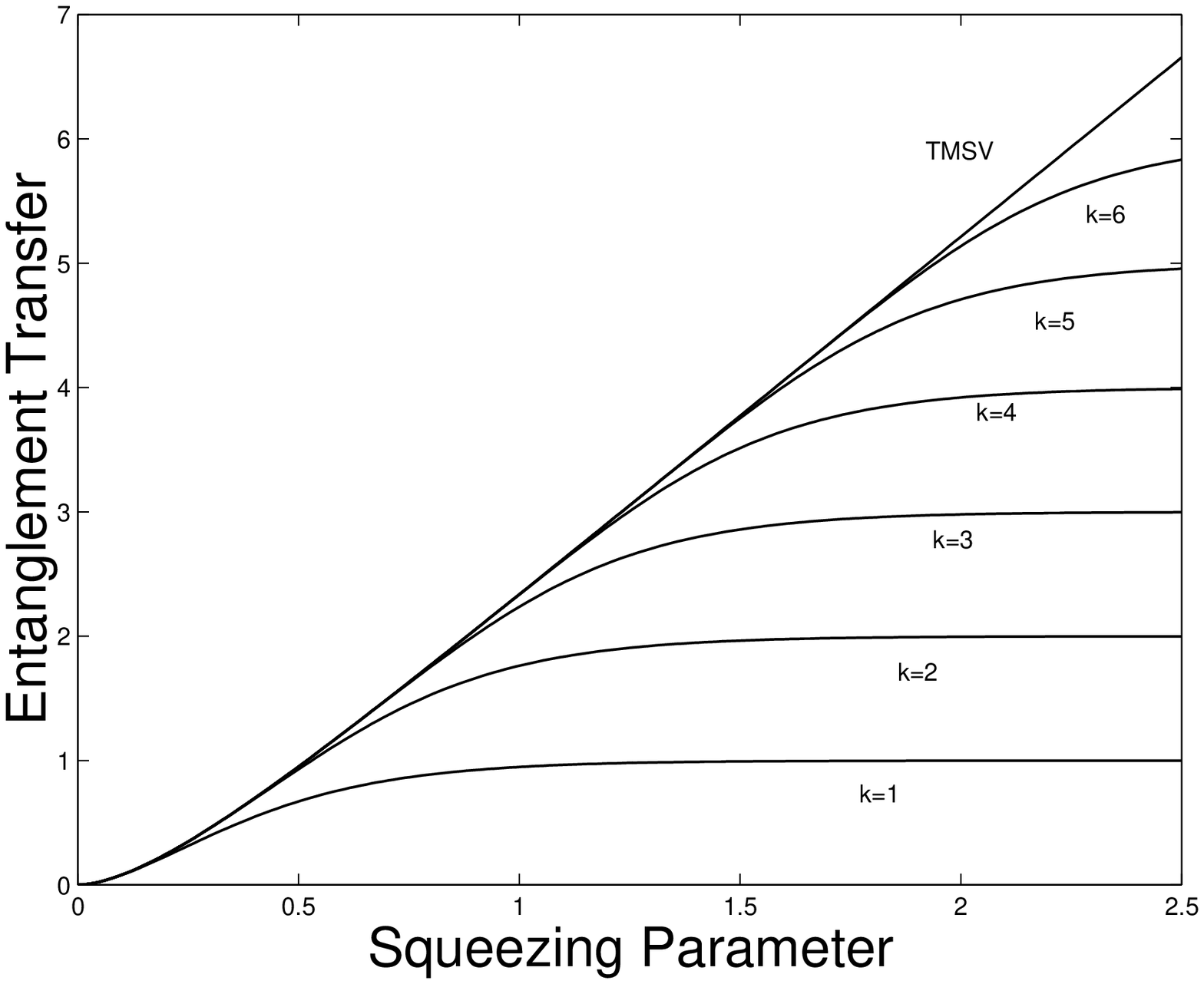}
\caption{.}
\end{figure}

Conversion of a digital number (a serial of bits) to an analog quantity has
the property that each bit is independent of other bits. No correlations
among these bits exist. That is the source is a discrete memoriless source.
D/A conversion is simply convert binary number to M-nary number. The
extension of the independence to quantum situation is that there are no
entanglements among the series of qubits. In the bipartite case, the state
before conversion will be a direct product of a series of two qubits. We
have the initial state $\left| \phi _1\right\rangle _{CD}\left| \phi
_2\right\rangle _{CD}$ $\cdots \left| \phi _k\right\rangle _{CD\text{ }}$,
where $\left| \phi _i\right\rangle _{CD\text{ }}=a_{00}^i\left|
--\right\rangle ^i+a_{01}^i\left| -+\right\rangle ^i+a_{10}^i\left|
+-\right\rangle ^i+a_{11}^i\left| ++\right\rangle ^i$. The process of
entanglement transfer is to transfer firstly the higher qubit pair ($k-th)$
to the CV bipartite state then the lower. The result of conversion will be $%
U_1^{+}(t_1)\left| \phi _1\right\rangle _{CD}U_2^{+}(t_2)\left| \phi
_2\right\rangle _{CD}\cdots U_k^{+}(t_k)\left| \phi _k\right\rangle _{CD%
\text{ }}\left| 00\right\rangle _{AB}=\left| \psi \right\rangle $ $%
_{AB}\prod_{i=1}^k\left| --\right\rangle ^i,$ where $\left| \psi
\right\rangle _{AB}=\sum_{n_1,\cdots n_k,m_1,\cdots
m_k=0}^1\prod_{j=1}^k(-1)^{m_{j+1}+n_{j+1}}i^{m_j+n_j}a_{n_jm_j}^j\left|
n_k\cdots n_1,m_k\cdots m_1\right\rangle ,$with $n=\sum_{j=1}^kn_j2^{j-1}$
denoted as $n_k\cdots n_1,$ $n_j=0,1.$ The Entanglement of the state $\left|
\psi \right\rangle _{AB}$ is equal to that of a state $\left| \psi ^{\prime
}\right\rangle =\sum_{n_1,m_1=0}^1i^{m_1+n_1}a_{n_1m_1}^1\left|
n_1,m_1\right\rangle
\prod_{j=2}^k(\sum_{n_j,m_j=0}^1(-i)^{m_j+n_j}a_{n_jm_j}^j\left|
n_j,m_j\right\rangle )$, thus it is equal to the sum of entanglements of
qunit pairs. We have $E(\left| \psi \right\rangle
_{AB})=\sum_{j=1}^kE(\left| \phi _j\right\rangle _{CD}).$ The conversion
procedure will convert a general qubit pair product state $\rho
_{1CD}\otimes \rho _{2CD}\otimes \cdots \otimes \rho _{kCD}$ into a
continuous variable state $\rho _{AB}$ while keeping the entanglement due to
local unitary operations.

Conversion of an analog quantity to a digital number has quantization error
due to the finite number of the destination bits. Speech signal is converted
to eight bits after sampling according to the standard protocol.. The
quantization error is small enough that it can not be sensed by ear. The
quantization of a quantum CV will also has quantization error if a finite
number of qubits is used. Some prior works have been done with the quantum
rate distortion theory\cite{Barnam} \cite{Devatak}. Let us consider quantum
A/D conversion of a single mode quantum state first. The initial CV state is
$\rho =\sum_{n,m=0}^\infty c_{nm}\left| n\right\rangle \left\langle m\right|
,$ the first step of conversion will be $\exp [-\frac i\hbar H_1t_1]\rho
\otimes \left| -\right\rangle \left\langle -\right| \exp [\frac i\hbar
H_1t_1]=$ $\sum_{n,m=0}^\infty (-1)^{m+n}\left| 2n\right\rangle \left\langle
2m\right| (c_{2n,2m}\left| -\right\rangle \left\langle -\right| +$ $%
ic_{2n,2m+1}\left| -\right\rangle \left\langle +\right| $ $%
-ic_{2n+1,2m}\left| +\right\rangle \left\langle -\right| $ $%
+c_{2n+1,2m+1}\left| +\right\rangle \left\langle +\right| )$. Then the
unitary transformation $\exp [-\frac i\hbar H_2t_2]$ is applied and so on,
at last $\exp [-\frac i\hbar H_kt_k]$ is applied. Each item of the CV part
will convert to a form of $\left| 2^kn\right\rangle \left\langle 2^km\right|
$. At this stage the entropy of the state remains intact. We obtain the
qubit series by tracing out the CV part and drop it. The tracing operation
will increase the total entropy by the triangle relation of the entropies.
The result $k-$qubit state usually has correlation among the qubits. The
correlation may even be considered as a kind of entanglement. Let us
consider the conversion of a CV state to two qubits, the result state after
tracing CV part will be $\sum_{k1,k2,l1,l2=0}^1i^{k1-k2-l1+l2}d_{k1k2l1l2}%
\left| k2,k1\right\rangle \left\langle l2,l1\right| $. Where $\left|
-\right\rangle ,\left| +\right\rangle $ are re-expressed by $\left|
0\right\rangle ,\left| 1\right\rangle $, $d_{k1k2l1l2}=%
\sum_mc_{4m+2k1+k2,4m+2l1+l2}.$ There may be entanglement between the first
and second qubits. For example, when the initial CV state is a coherent
state $\left| \alpha \right\rangle $ with real parameter $\alpha $, the
entanglement can be quite high. The concurrence increases from 0 ($\alpha =0$%
) to 0.9462 ($\alpha =1.29$) , then decreases to 0.8271($\alpha =1.92$),
after that it monotonically increases and at $\alpha =3.4$ revives to the
first maximum. Thus quantum A/D conversion can produce entanglement with
quite high quality. The entanglement produced from coherent state relies on
the phase angle of the complex parameter, for a coherent state with pure
imaginary parameter, there is no converted entanglement at all between the
two qubits.

An example of totally memoriless qubits series can be obtained from
converting the thermal state. The conversion fidelity can be considered at
the basis of CV and qubit series as well. We can convert the $k-$qubit
series back to CV state and we obtain a state $\rho ^{\prime }$, the
conversion fidelity $F$ is the fidelity between $\rho $ and $\rho ^{\prime }$%
. The distortion is $D=1-F.$ The rate distortion problem is that what is the
minimal coding rate $R$ for a given source under the constrain of distortion
$D.$ Thermal state source $\rho =(1-v)\sum_{n=0}^\infty v^n\left|
n\right\rangle \left\langle n\right| $ is one of the simplest source. $\rho
^{\prime }=\frac{(1-v)}{1-v^{2^k}}\sum_{n=0}^{2^k-1}v^n\left| n\right\rangle
\left\langle n\right| $. The distortion will be $D=1-\sqrt{1-v^{2^k}},$ Thus
$k=\log [\log [1-(1-D)^2]/\log v].$


\begin{thebibliography}{9}
\bibitem{Barnum}  H. Barnum, Quantum rate-distortion coding, Phys. Rev. A
62, 42309(2000).

\bibitem{Devetak}   I. Devetak and T. Berger,  Quantum rate-distortion
theory for memoryless sources. IEEE Transactions on Information Theory
48(6): 1580-1589 (2002).
\end{thebibliography}
\end{document}